\documentclass[trackchanges, twocolumn, twocolappendix]{aastex631}
\usepackage{CJK}
\usepackage{mathptmx}
\usepackage{amsmath}
\usepackage{ulem}
\hypersetup{citecolor=[RGB]{0,80,150}}

\newcommand\hi{{\rm H}{\textsc i}}
\newcommand{\atlas}{\textsc{Atlas}$^{\rm 3D}$}

\defcitealias{Yu2022a}{Y22}
\defcitealias{Ellison2025}{E25}

\shorttitle{\hi\ spectral shape of PSBs}
\shortauthors{Huang et al.}

\graphicspath{{./}{figures/}}

\begin{document}
\begin{CJK*}{UTF8}{gbsn}

\title{Decoding the Single-peaked \textsc{Hi} Spectra of Low Redshift Post-starburst Galaxies}

\correspondingauthor{Jing Wang}
\email{jwang\_astro@pku.edu.cn}

\author[0000-0003-2863-9837]{Qifeng Huang (黄齐丰)}
\affiliation{Kavli Institute for Astronomy and Astrophysics, Peking University, Beijing 100871, China}

\author[0000-0002-6593-8820]{Jing Wang (王菁)}
\affiliation{Kavli Institute for Astronomy and Astrophysics, Peking University, Beijing 100871, China}

\author[0000-0002-1768-1899]{Sara L. Ellison}
\affiliation{Department of Physics \& Astronomy, University of Victoria, Finnerty Road, Victoria, BC V8P 1A1, Canada}

\author[0009-0007-9363-2582]{Zezhong Liang (梁泽众)}
\affiliation{Kavli Institute for Astronomy and Astrophysics, Peking University, Beijing 100871, China}

\author[0000-0002-4250-2709]{Xuchen Lin (林旭辰)}
\affiliation{Kavli Institute for Astronomy and Astrophysics, Peking University, Beijing 100871, China}

\author[0000-0002-5679-3447]{Dong Yang (杨冬)}
\affiliation{Kavli Institute for Astronomy and Astrophysics, Peking University, Beijing 100871, China}

\begin{abstract}

Recent observations with the Five-hundred-meter Aperture Spherical Telescope (FAST) have revealed abundant reservoirs of neutral hydrogen (\hi) in low redshift post-starburst galaxies (PSBs), raising the question of why star formation ceases rapidly in these systems. In this study, we present a detailed analysis of the shape of the integrated \hi\ spectra of 67 PSBs. We find that PSBs exhibit significantly higher \hi\ spectral concentration values ($K$) compared to a matched sample from xGASS, and are more comparable to those of starburst galaxies. By extending our analysis to spatially resolved \hi\ data from THINGS and \atlas, we show that both centrally concentrated \hi\ distributions and dynamically unsettled \hi\ can effectively increase $K$, while non-axisymmetric structures only contribute to the scatter of the $K$ distribution. Distinguishing between central concentration and dynamically unsettled gas as the origin of high $K$ can be achieved by measuring the spectral asymmetry ($A_{\rm F}$), making the $K$-$A_{\rm F}$ plane a powerful diagnostic tool for identifying galaxies with unsettled \hi\ using integrated spectra alone. Based on their location in the $K$-$A_{\rm F}$ plane, we find that most PSBs are not dominated by unsettled \hi, but rather exhibit elevated central gas concentration. Both modes of gas redistribution in PSBs may eventually contribute to their quenching.

\end{abstract}


\section{Introduction} \label{sec:intro}

Post-starburst galaxies (PSBs) represent a unique galaxy population that have recently experienced a starburst phase, followed by a rapid decline in star formation rates (SFRs). This evolutionary track leaves distinctive features in their optical spectra \citep{Dressler1983, Poggianti1999, Wild2007}. Due to their unique star formation histories (SFHs) and structural similarities with early-type galaxies (ETGs), low-redshift PSBs are widely regarded as systems caught in the act of quenching---transitioning from the star-forming main sequence to the quiescent state (e.g., \citealt{Yang2008, Chen2022}; but see \citealt{Pawlik2018,Pawlik2019}). As such, they serve as critical laboratories for understanding the bimodality of galaxy populations and the physical mechanisms driving galaxy quenching (see \citealt{French2021} for a review).

As the fuel for star formation, the interstellar medium (ISM) of PSBs holds key clues to the rapid cessation of star formation in these systems. Previous observational studies have consistently revealed that PSBs retain significant molecular gas reservoirs, which display a suppressed efficiency in transitioning into dense gas and forming stars \citep{Rowlands2015, Alatalo2016, French2018a}, with simulations showing similar trends \citep{Davis2019a}. In cases where high-resolution observations are available, the molecular gas shows extremely high turbulent pressure, potentially accounting for the low star formation efficiency (SFE) observed in PSBs \citep{Otter2022,Smercina2022}. By tracking the time evolution following the starburst, it has also been found that the mass budget of star-forming gas in PSBs declines over time \citep{Rowlands2015, French2018}. However, this gas depletion cannot be fully explained by either residual star formation \citep{French2018} or gas outflows \citep{Alatalo2015, Fodor2025}. Instead, transitioning from the molecular phase to warmer phases of the ISM, including the \hi, is likely necessary to reconcile the current observations.

Studies of PSBs using \hi\ have historically been limited by small sample sizes, shallow detection limits, or atypical environments \citep{Chang2001, Buyle2006, Zwaan2013}. It was only recently that \citet[][hereafter \citetalias{Ellison2025}]{Ellison2025} provided a comprehensive demographic study of \hi\ content in low-redshift PSBs. By analyzing a mass-complete sample of 68 PSBs, they found significant \hi\ reservoirs, with \hi\ fractions intermediate between those of star-forming galaxies and galaxies in the green valley. 
The plentiful \hi\ left in PSBs therefore leaves open the possibility of re-igniting future star formation. However, this possibility depends critically on the state (e.g., its dynamics) and location (e.g., still in the disk or expelled towards the halo) of \hi.

In principle, the best assessment of the status of the remaining \hi\ in PSBs is via detailed spatial mapping, but unfortunately, spatially resolved \hi\ maps are expensive to obtain.
On the other hand, as large extragalactic surveys continue to expand, our samples of integrated (or barely resolved) \hi\ spectra are expanding rapidly (e.g., FASHI; \citealt{Zhang2024}).
Thus, extracting meaningful information from \hi\ spectra becomes essential to probe \hi\ properties in large samples and possibly at cosmic distances \citep{Wang2020, Yu2022a, Peng2023}. In this paper, we focus on the concentration (i.e., ``peakedness'' or kurtosis) of \hi\ spectra following the definition of \citet[][hereafter \citetalias{Yu2022a}]{Yu2022a} as a potential tracer of \hi\ spatial distribution and kinematics. Using public interferometry data from The \hi\ Nearby Galaxy Survey \citep[THINGS;][]{Walter2008} and the \atlas\ project \citep{Cappellari2011a, Serra2012}, we evaluate how the spatial distribution of \hi\ influences the concentration of integrated spectra and use this information to interpret the observed spectral shape of PSBs. Through this analysis, we aim to constrain the \hi\ properties of PSBs and distinguish between possible scenarios for their quenching.

This paper is organized as follows. Section \ref{sec:data} describes the data used in this paper and introduces the measurement of spectral concentration of \hi. In Section \ref{sec:result}, we compare the spectral concentration of PSBs with other galaxy samples. We examine the drivers of spectral concentration and discuss the implications of \hi\ properties in PSBs in Section \ref{sec:discussion}. Finally, we summarize the paper and provide future perspectives in Section \ref{sec:conclusion}.

\section{Data and Analysis} \label{sec:data}

\subsection{A mass-complete sample of PSBs} \label{subsec:sample}

The PSB sample used in this paper is taken from \citetalias{Ellison2025}. Briefly, they combine traditional E+A PSBs \citep{Goto2005} and a broader PSB sample selected with the PCA method \citep{Wild2007, Wilkinson2022} from the 7th Data Release of Sloan Digital Sky Survey (SDSS; \citealt{Abazajian2009}). The sample is further limited to a stellar mass range of $\log M_\ast/M_\odot>9.5$ and redshift $0.01<z<0.04$, yielding 86 PSBs after excluding sources with significant confusion in \hi\ observations. Clean \hi\ spectra are obtained for 43 PSBs at $z<0.0364$ using FAST \citep{Nan2011}, which targets a detection limit similar to the extended GALEX Arecibo SDSS Survey \citep[xGASS;][]{Catinella2010, Catinella2018}, which can be used as a control sample. For the remaining PSBs, we utilize archival data, including 17 \hi\ spectra from ALFALFA \citep{Giovanelli2005}, 2 from xGASS, and 5 from \hi-MaNGA \citep{Masters2019, Stark2021}. The final sample consists of 67 PSBs with available integrated \hi\ spectra. More details on the sample selection, FAST observations, and data reduction can be found in \citetalias{Ellison2025}. We note that removing PSBs with \hi\ confusion (25 out of 111; \citetalias{Ellison2025}) may bias the sample against systems undergoing strong interactions, which can disturb the \hi. However, since galaxies with \hi\ confusion are also removed from the comparison sample (Section \ref{subsec:control}), and PSBs are not preferentially found in close pairs at these separations \citep{Ellison2022}, the resulting sample bias is minimized.

For the PSB sample, we use stellar masses from the MPA-JHU catalogs \citep{Kauffmann2003a}. We cross-match the sample with the NASA-Sloan Atlas \citep{Blanton2011} to obtain measurements of optical concentration ($C_r$), defined as the ratio between the radii enclosing 90\% and 50\% of the Petrosian flux, and apparent axis ratio $q\equiv b/a$, both measured in $r$-band. The inclination angles ($i$) are estimated using the following equation \citep{Hubble1926}, incorporating a stellar mass-dependent intrinsic disk thickness ($q_0$) from \citet{Sanchez-Janssen2010}:
\begin{equation}
    \cos^2{i} = \frac{q^2-q_0^2(M_\ast)}{1-q_0^2(M_\ast)}. \label{eq:i}
\end{equation}
This method tends to underestimate the inclination angles for spheroid-dominated systems, as is likely the case for some PSBs in our sample with $C_r>2.6$ \citep{Strateva2001}. Correcting for this bias would further enhance the significance of our results, as we will see in Section \ref{sec:result}.

The same catalogs and methods are used to obtain properties of the comparison sample introduced in the next section. 

\begin{figure*}
    \centering
    \includegraphics[width=\linewidth]{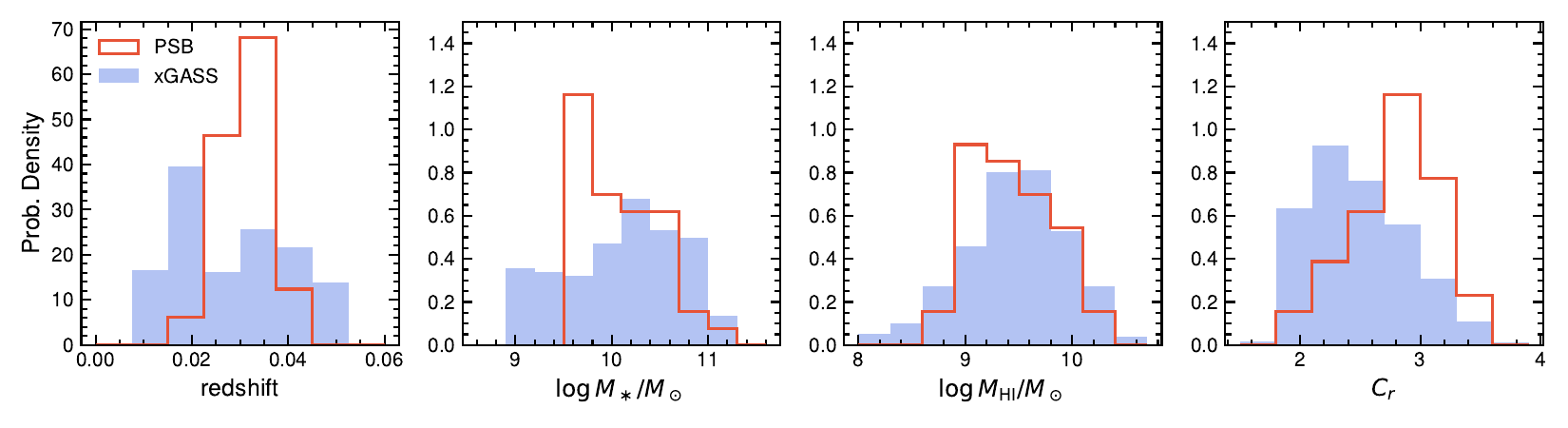}
    \caption{Normalized distributions of sample properties for PSBs (red open) and the comparison sample from xGASS (blue filled).}
    \label{fig:sample}
\end{figure*}

\subsection{The comparison sample} \label{subsec:control}

xGASS provides a representative sample of low-redshift ($0.01 <z<0.05$) galaxies with deep, uniform \hi\ observations \citep{Catinella2018}. It serves as a benchmark for evaluating the unique properties of PSBs, given our matched \hi\ detection limits. 
To construct a sample for comparison with the PSBs, we select \hi-detected galaxies from xGASS that do not have a confusion flag (HI\_FLAG $\leq2$) and are not contaminated by radio frequency interference (\citetalias{Yu2022a}). To uniformly remove the \hi\ non-detections in both PSBs and xGASS galaxies, we adopt the updated detection threshold from \citetalias{Ellison2025}: $M_{\rm HI}/M_\ast>0.02$ for galaxies with $\log M_\ast/M_\odot>10.55$, and $\log M_{\rm HI}/M_\ast>8.85$ at lower stellar masses. This more conservative threshold also excludes marginal \hi\ detections, where the spectral shape cannot be reliably measured. Our final sample includes 43 PSBs and 630 xGASS galaxies that meet these criteria.

Figure \ref{fig:sample} presents the normalized distribution of redshift, stellar mass, \hi\ mass, and optical concentration of PSBs and xGASS galaxies. We can see that PSBs occupy a similar parameter space with xGASS, except that they are more likely to have higher optical concentrations. In Section \ref{sec:result}, we will control for key galaxy properties for a more robust comparison of the two samples. Our focus is on the concentration of their \hi\ spectra, which is closely related to the spatial distribution of \hi, as will be shown later.

\subsection{Concentration of integrated \hi\ spectra} \label{subsec:K}

\begin{figure}
    \centering
    \includegraphics[width=\linewidth]{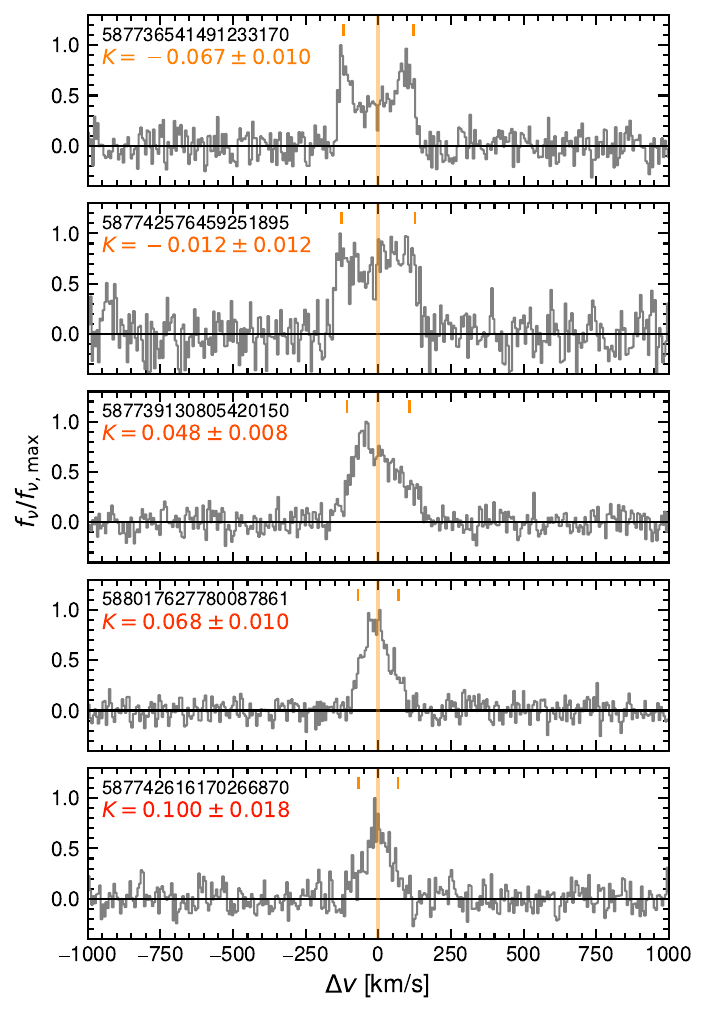}
    \caption{Examples of PSB spectra obtained with FAST, ordered by increasing $K$ and normalized by the peak flux density. Each panel displays the $K$ value and SDSS DR7 objID in the upper left corner. The short orange bars at the top of each panel represent the velocity range of $\Delta v= \pm V_{85}$, where $V_{85}$ is the half width enclosing 85\% of the total flux, as defined for $K$.
    }
    \label{fig:K-rank}
\end{figure}
 
The shape of integrated \hi\ spectra can be broadly and qualitatively classified into three categories: double-peaked, flat-topped, and single-peaked. In order to have a more quantitative measure, we use the nonparametric quantity $K$ introduced by \citetalias{Yu2022a} to quantify the concentration of \hi\ spectra. The definition of $K$ is the integrated area between the normalized growth curve---normalized to the point where the flux reaches 85\% of the total flux---and the diagonal line of unity (see Figure 2 of \citetalias{Yu2022a}). A negative $K$ indicates a double-horned \hi\ spectrum, while a positive value corresponds to a single-peaked spectrum. Flat-topped spectra are characterized by $K\approx0$. For reference, a Gaussian line profile has $K=0.079$. A 2-D \hi\ disk with a uniform surface density, a flat rotation curve, and no velocity dispersion has $K=-0.092$. Figure \ref{fig:K-rank} provides some examples of FAST spectra ordered by $K$. It is clear that the spectra change from double-horned to single-peaked as $K$ increases (from top to bottom panels in Fig. \ref{fig:K-rank}). The physical implications of $K$ will be explored in Section \ref{sec:discussion}.

We measure the values of $K$ for all galaxies used in this study and apply the S/N-dependent correction for systematic biases following \citetalias{Yu2022a}. The associated uncertainties ($\sigma_K$) are estimated by adding random Gaussian noise to the original spectra, which increases the noise level by a factor of $\sim\sqrt{2}$ and makes $\sigma_K$ a conservative estimate. To minimize systematics between different surveys, we resample the spectra to a uniform channel width of $\rm 5.5~km~s^{-1}$. For galaxies with ALFALFA spectra, we adopt the $K$ values reported by \citetalias{Yu2022a}. To ensure consistency in the measurement procedures, we compare our measurements with those of \citetalias{Yu2022a} for 1000 randomly selected galaxies in their sample, finding negligible systematic offsets ($0.001\pm0.004$). The $K$ values of the PSB sample are provided in Appendix \ref{app:tab}.

\section{Single-peaked \hi\ spectral shape of PSBs} \label{sec:result}

\begin{figure*}
    \centering
    \includegraphics[width=0.9\linewidth]{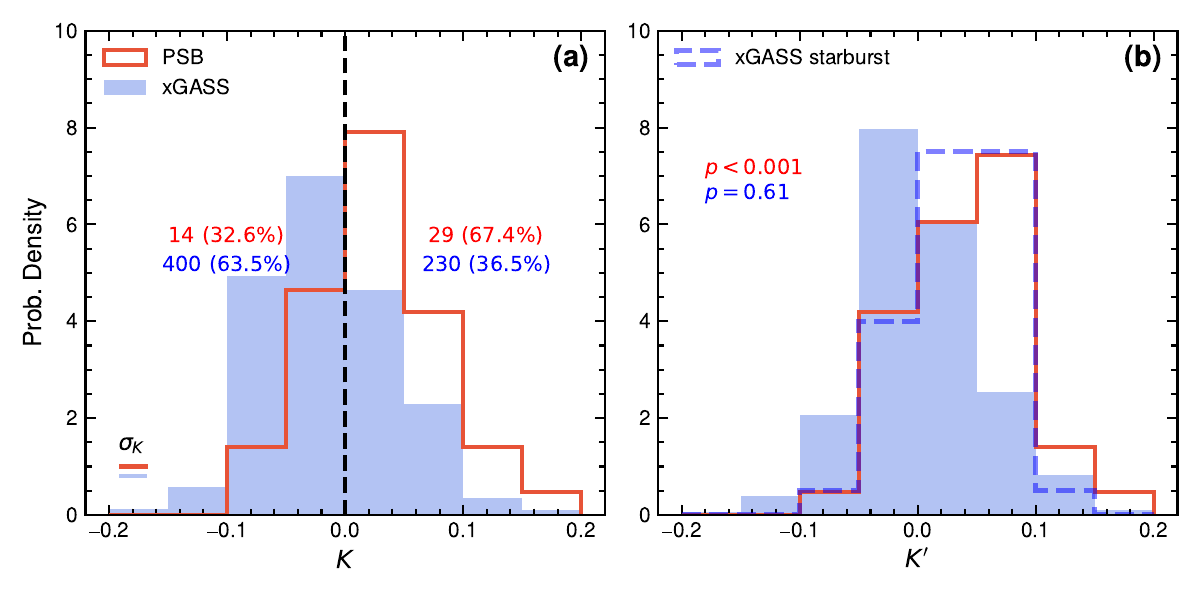}
    \caption{\textbf{(a)} Normalized distribution of $K$. The same color scheme is used as Figure \ref{fig:sample}. The short lines in the lower left represent the typical uncertainty in $K$ ($\sigma_K$) for each sample. The vertical dashed line marks $K=0$, the division of single-peaked and double-horned \hi\ spectral shapes. The numbers (percentages) of galaxies in both categories are also provided. \textbf{(b)} Normalized distribution for $K'$ for PSBs, xGASS, and starburst galaxies in xGASS. The $p$-value in red is evaluated between PSBs and xGASS, while the blue one is calculated between PSBs and the starburst galaxies.}
    \label{fig:K-pdf}
\end{figure*}

In Figure \ref{fig:K-pdf}a, we compare the $K$ distributions among the samples defined in Section \ref{sec:data}. The offset between PSBs and the xGASS sample clearly shows the peculiarity of PSBs as a special population. While $36.5\%\pm2.6\%$ galaxies in xGASS exhibit single-peaked \hi\ spectra with $K > 0$, this fraction for the PSBs is almost double ($67.4\%\pm13.3\%$). Thus, we conclude that most of the low-redshift PSBs in our sample exhibit single-peaked \hi\ spectra. 

When relating the spectral concentration $K$ to the spatial distribution of \hi, it is important to note that $K$ is influenced by two additional factors besides the \hi\ spatial concentration \citep[e.g.,][]{Yu2022}. First, the relative extent of the rising part of the rotation curve and the \hi\ disk affects the fraction of \hi\ near the central velocity. Second, at lower inclinations, the effect of dispersion-dominated processes (e.g., turbulence and thermal motion) on the spectral shape becomes more significant relative to rotation, leading to a negative correlation between $K$ and the inclination \citep{El-Badry2018}. To ensure a more adequate comparison, we control the related properties ($M_\ast$, $M_{\rm HI}$, $C_r$, and $i$; \citealt{Yu2022}) between PSBs and xGASS. As the sample size is limited, a one-to-one matching is not feasible to control four parameters simultaneously. Instead, we subtract the parameter dependence of $K$ by fitting a hyperplane between $\theta=(\log M_\ast/M_\odot$, $\log M_{\rm HI}/M_\odot$, $C_r$, $\sin i)$ and $K$ using xGASS:
\begin{equation}
    K_0(\theta) = \alpha \theta^{\rm T}+\beta. \label{eq:K0}
\end{equation}
Using a Huber regressor that is robust to outliers,\footnote{\url{https://scikit-learn.org/}}  we obtain the best-fit parameters $\alpha=(-0.0094$, $-0.0256$, $0.0042$, $-0.0596)$ and $\beta=0.352$. The $K$-excess, denoted as $K'$, is derived by subtracting $K_0$ from the original $K$ and provides a fair comparison of \hi\ spectral concentration among different galaxies. By definition, we expect $K'\approx0$ for a typical galaxy in the xGASS sample. We find no residual dependence of $K'$ on the aforementioned parameters (Appendix \ref{app:k-dep}). 
\footnote{Given the correlation between $K$ and $\sin i$, an uncertainty of $10^\circ$ in inclination would result in a change of only $<0.01$ in $K'$. Therefore, underestimating the inclination angles of PSBs (Section \ref{subsec:sample}) does not affect our results, as the median $K'$ would change by no more than 0.015. This upper limit is obtained by considering the extreme case where all PSBs are edge-on, i.e., where $i=90^\circ$.}

The distributions of $K'$ are shown in Figure \ref{fig:K-pdf}b, which confirms that PSBs have a higher spectral concentration than the general galaxy population in xGASS. The $p$-value from a Kolmogorov--Smirnov (K-S) test is smaller than 0.001, indicating a very significant elevation of \hi\ spectral concentration in PSBs. 
Figure \ref{fig:K-pdf}b also compares the PSBs with 40 starburst galaxies in xGASS, defined as those lying $>0.4$ dex above the star-forming main sequence fitted by \citet{Saintonge2016}. The SFRs are obtained by fitting the far-ultraviolet to mid-infrared broad-band SEDs \citep{Salim2018}. No significant difference is observed between PSBs and starbursts ($p>0.1$), consistent with previous findings that higher $K$ is usually associated with enhanced star formation activity at a given $M_\ast$ (\citealt{Yu2022}), likely driven by centrally concentrated \hi\ \citep{Wang2020}. This suggests that when galaxies transit from the starburst phase to the PSB phase, the \hi\ distribution and kinematics are retained.

\section{Discussion} \label{sec:discussion}

In the previous section, we demonstrated that PSBs have $K$ (and $K'$) values whose single-peaked nature is more similar to on-going starbursts than ``normal'' xGASS galaxies. In this section, we investigate the primary drivers of $K$ using spatially resolved \hi\ data (Section \ref{subsec:k-prof} to \ref{subsec:k-diff}), and discuss the implications of our results on the evolution of PSBs (Section \ref{subsec:k-evo}).

\subsection{Spatial concentration of \hi} \label{subsec:k-prof}

\begin{figure*}
    \centering
    \includegraphics[width=\linewidth]{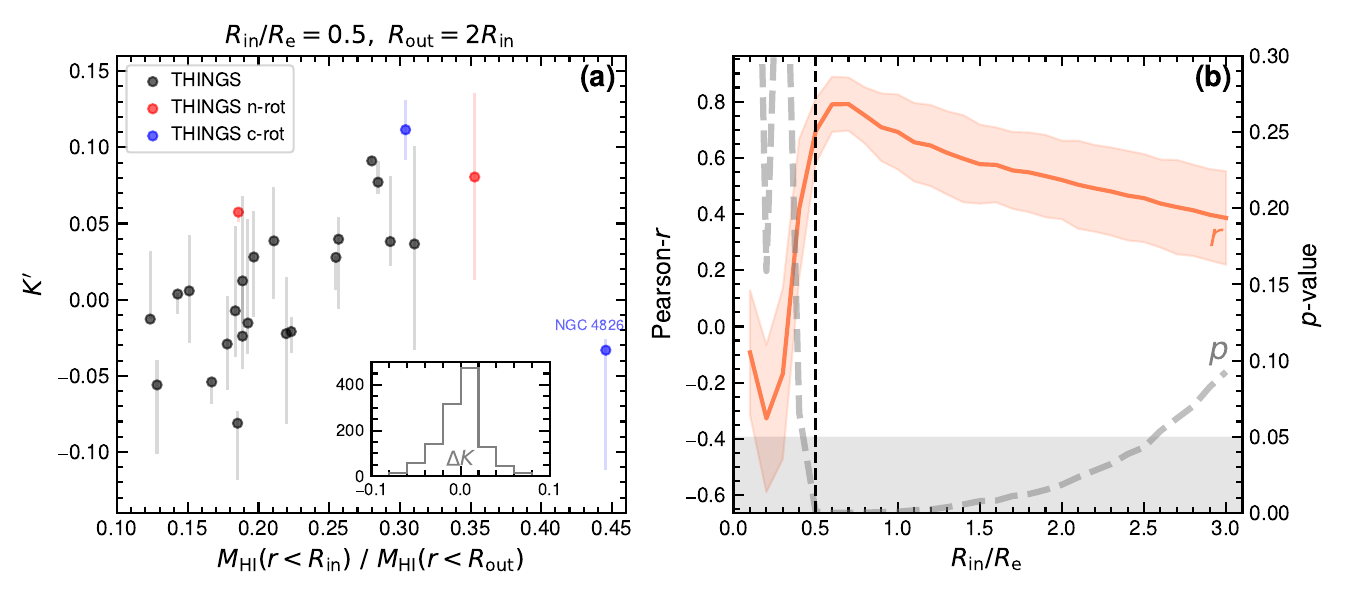}
    \caption{\textbf{(a)} Correlation between $K'$ and the spatial \hi\ concentration for 25 THINGS galaxies. The red and blue dots represent galaxies with no obvious rotation pattern (n-rot) and with counter-rotating \hi\ disks (c-rot), respectively. The error bars and the inset panel display the change in $K$ when artificially rotating the \hi\ mass distribution within the disk plane (see Section \ref{subsec:k-naxis} and Appendix \ref{app:int} for details). \textbf{(b)} The Pearson-$r$ correlation coefficients and $p$-values as a function of $R_{\rm in}$. The vertical dashed line indicates the radius where panel (a) is plotted. The shaded region at the bottom corresponds to $p < 0.05$. The uncertainties in $r$ are obtained from 1000 bootstraps. The correlation is strongest at $R_{\rm in}\sim 0.7R_{\rm e}$ and becomes insignificant at $R_{\rm in} \gtrsim 2.5R_{\rm e}$.}
    \label{fig:k-prof}
\end{figure*}

In regularly rotating, dynamically cold \hi\ disks, the spatial concentration of \hi\ is regarded as a critical driver of spectral concentration \citep{Yu2022}. Although this assumption is intuitive, it has not yet been rigorously tested using spatially resolved \hi\ maps from observations. Here we measure the \hi\ concentration of galaxies from the THINGS survey, which offers high spatial resolution and enables a detailed description of spatial \hi\ distributions \citep{Walter2008}. We use elliptical apertures in the moment-0 maps to calculate the ratio of \hi\ masses within an inner radius ($R_{\rm in}$) and an outer radius that is twice the size of the inner one ($R_{\rm out}=2R_{\rm in}$). We then measure the $K$ values from the integrated spectra of each galaxy and correct for the inclination angle dependence as in Section \ref{sec:result}. Applying inclination measurements from kinematic modeling does not affect our results \citep[e.g.,][]{deBlok2008}.

Figure \ref{fig:k-prof}a illustrates the relationship between spectral concentration ($K'$) and spatial concentration of \hi\ measured at $R_{\rm in}=0.5R_{\rm e}$, where $R_{\rm e}$ is the effective radius measured at 3.6 $\mu{\rm m}$ wavelength \citep{Salo2015, Sanchez-Alarcon2025}. Nine THINGS galaxies without a measurement of $R_{\rm e}$ are excluded in the figure. The Pearson-$r$ parameter ($r=0.69\pm0.12$) indicates a strong positive correlation between these quantities, demonstrating that $K'$ effectively traces the spatial concentration of \hi\ in THINGS galaxies.\footnote{Galaxies with no rotation pattern in the moment-1 maps (n-rot) and with counter-rotating \hi\ disks (c-rot) are not included in the correlation tests. NGC 4826 deviates significantly from this correlation due to its compact counter-rotating inner disk.}
To further investigate the radial range to which $K'$ is the most sensitive, we plot the Pearson-$r$ coefficient and the $p$-value as a function of $R_{\rm in}$ in Figure \ref{fig:k-prof}b. We find that the strongest correlation occurs at $R_{\rm in}\sim 0.7R_{\rm e}$, while the correlation becomes insignificant at $R_{\rm in} \gtrsim 2.5R_{\rm e}$. This result suggests that $K'$ is mainly affected by the \hi\ concentration in the inner disk at $R_{\rm in} < R_{\rm e}$, consistent with the fast-rising rotation curves for typical THINGS galaxies \citep{deBlok2008}.

\subsection{Non-axisymmetric structures} \label{subsec:k-naxis}

When observing a rotating galactic disk, regions near the systematic velocity are not limited to the galaxy center but can also appear along the kinematic minor axis and in areas with low radial velocity, often produced by non-axisymmetric structures such as spiral arms and bars. As a result, the spectral concentration $K$ is expected to be altered by the azimuthal inhomogeneity in the \hi\ distribution within the galactic disk. 

\begin{figure*}
    \centering
    \includegraphics[width=\linewidth]{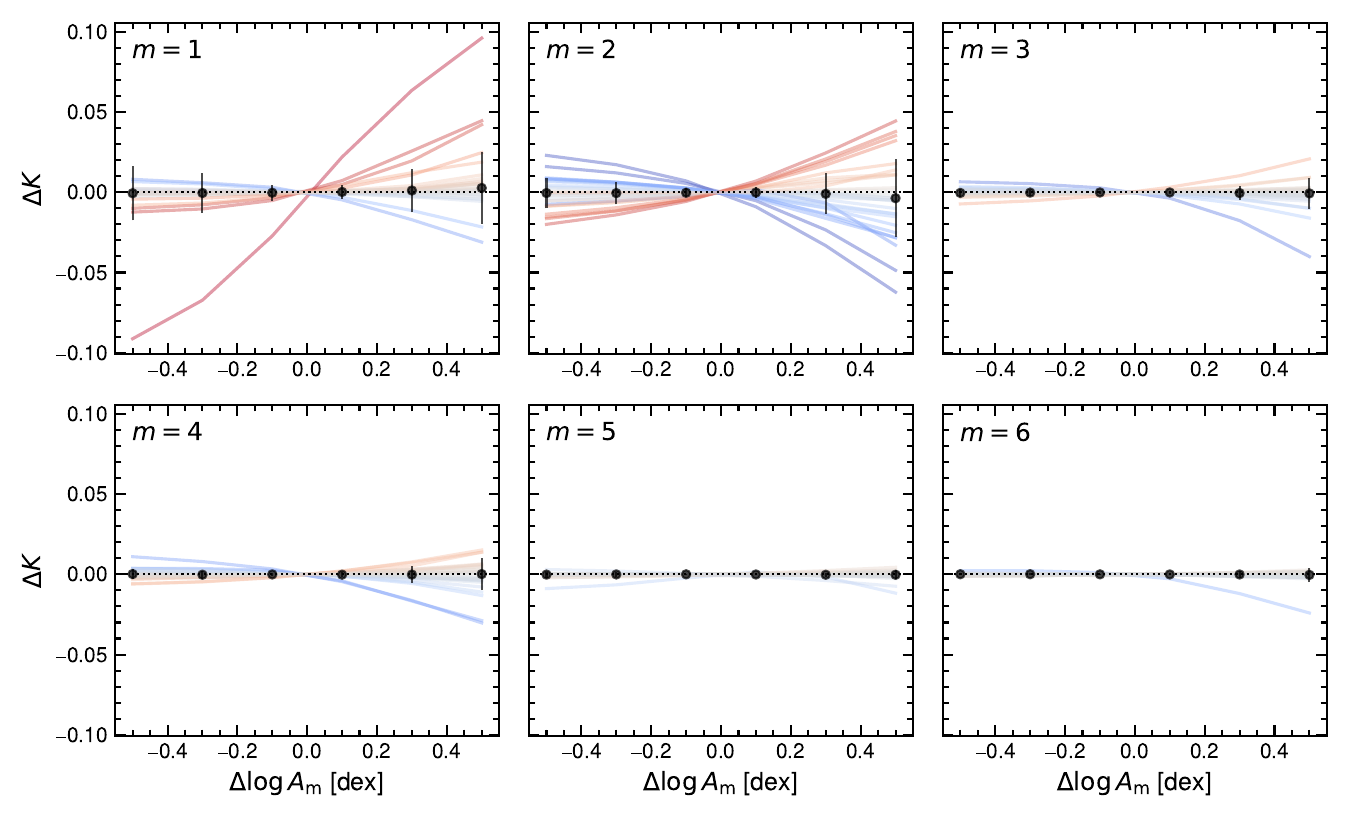}
    \caption{Variations in spectral concentration ($\Delta K$) as a function of changes in Fourier amplitudes ($\Delta \log A_{\rm m}$) for modes $m=1$ to $m=6$. Solid lines represent individual THINGS galaxies, color-coded by the $\Delta K$ value at $\Delta \log A_{\rm m}=0.5$. Black dots with error bars denote the median values and standard deviations of $\Delta K$ for all 34 THINGS galaxies.}
    \label{fig:k-naxis}
\end{figure*}

To mimic random azimuthal variations in \hi\ surface density, we generate mock \hi\ spectra by rotating the THINGS \hi\ moment-0 maps within the disk mid-plane while keeping the moment-1 and moment-2 maps unchanged (see Appendix \ref{app:int} for details). We then measure the $K$ values for these mock spectra and calculate the offsets from the original $K$, denoted as $\Delta K$. In such cases, the amplitudes of the azimuthal variations are inherited directly from realistic galaxies and are representative of galaxies with diverse late-type morphologies, as designed by THINGS \citep{Walter2008}. In Figure \ref{fig:k-prof}a, the maximum ranges of $\Delta K$ for individual THINGS galaxies are indicated as error bars, and the distribution of $\Delta K$ is displayed in the inset panel. The mean and standard deviation of $\Delta K$ are $-3.5\times10^{-5}$ and 0.024, respectively. These values indicate that while azimuthal variations in \hi\ do not systematically shift the average $K$ for galaxy samples, they contribute to the scatter in $K$ distributions (which is 0.055 for xGASS and 0.053 for PSBs), consistent with the random orientations of real galaxies. 

To further evaluate the contribution of different non-axisymmetric components, we manually vary the amplitudes of different Fourier modes within a range of $-0.5$ to $0.5$ dex for each THINGS galaxy (see Appendix \ref{app:int} for details) and see how $K$ changes accordingly. Figure \ref{fig:k-naxis} illustrates the dependence of $\Delta K$ on the amplitude variations in the six leading Fourier modes. The results show that $K$ is mostly affected by the $m=2$ (dipole) and $m=1$ (lopsidedness) modes. This behavior is expected because $K$ can be boosted (decreased) by piling up \hi\ along the kinematic minor (major) axis while removing it in the orthogonal direction. Fourier modes with $m\geq5$ exhibit negligible effects on $K$. Moreover, the median $\Delta K$ (black dots in Figure \ref{fig:k-naxis}) for all Fourier modes is consistent with zero, reaffirming that non-axisymmetric structures mainly contribute to the scatter of $K$ rather than systematically altering its value. These results tell us that we can hardly infer the radial distribution of \hi\ for any individual galaxies from $K$, unless the azimuthal variations are properly accounted for.

\subsection{Unsettled \hi\ distribution} \label{subsec:k-a}

In previous sections, we assumed that \hi\ in galaxies is located in a rotating disk, which is not always the case. To extend the THINGS-based analysis to ETGs and consider a broader range of \hi\ morphologies and kinematics, we include the \atlas\ sample and measure the concentration $K$ of their integrated \hi\ spectra (see Appendix \ref{app:int} for details). 
Besides, we also utilize the spectral asymmetry for all the samples ($A_{\rm F}$; e.g., \citealt{Bok2019, Deg2020}; \citetalias{Yu2022a}) in the following analysis. $A_{\rm F}$ is defined as the ratio between the \hi\ flux on one side of the spectral center and the other, requiring $A_{\rm F}\geq1$.

\begin{figure*}
    \centering
    \includegraphics[width=\linewidth]{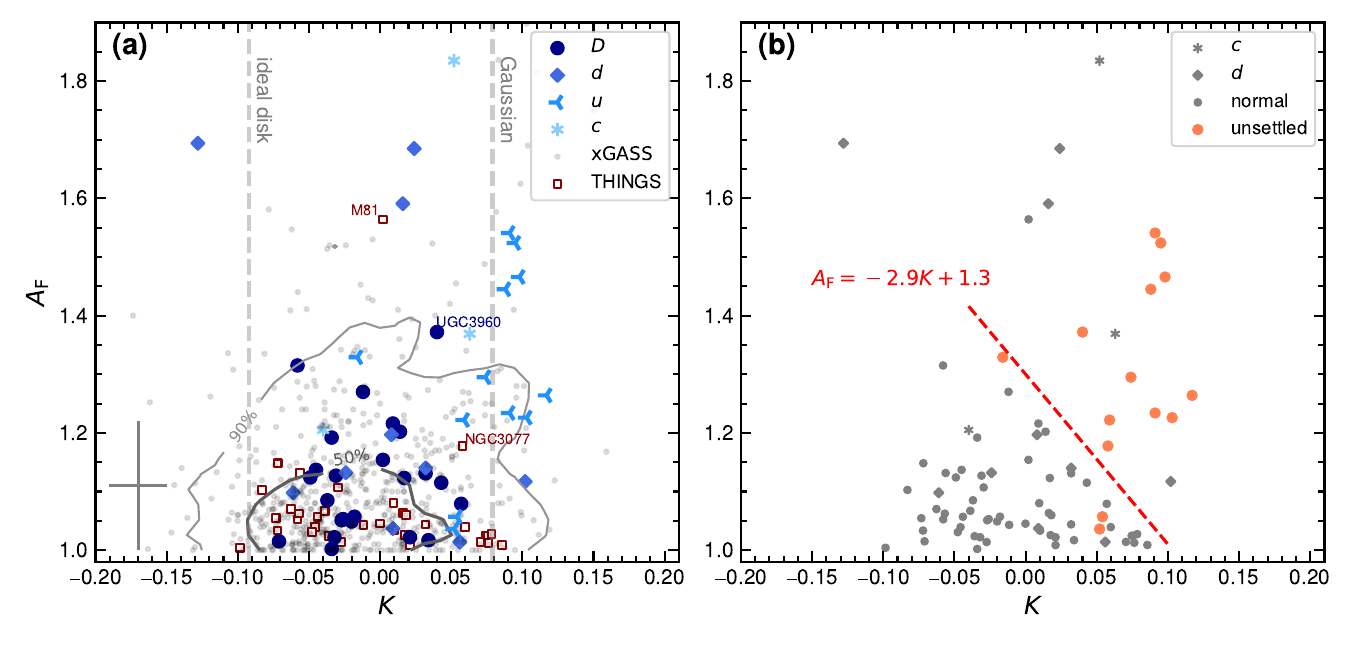}
    \caption{\textbf{(a)} Distribution of galaxies in the $K$-$A_{\rm F}$ plane. Galaxies from \atlas, xGASS, and THINGS are shown as blue symbols, gray dots, and dark red squares, respectively. The \atlas\ symbols follow \citet{Serra2012}: $D$ = regularly rotating \hi\ disks larger than the stellar disk, $d$ = regularly rotating \hi\ disks smaller than the stellar disk, $u$ = \hi\ with unsettled morphology/kinematics, and $c$ = \hi\ in small, scattered clouds. The vertical dashed lines mark the $K$ values for an idealized disk and a Gaussian profile (Section \ref{subsec:K}), respectively. Contours enclosing 50\% and 90\% xGASS galaxies are shown in gray. The typical measurement uncertainty of xGASS is provided in the lower left. \textbf{(b)} \atlas\ and THINGS galaxies in the $K$-$A_{\rm F}$ plane. The unsettled \hi\ disks are highlighted in red, including all the $u$-types, UGC 3960, and NGC 3077. The remaining galaxies are shown in gray. The red dashed line marks an empirical boundary separating normal ($D$-types and THINGS) and unsettled \hi\ disks (Equation \ref{eq:k-a}).
    \label{fig:k-a}}
\end{figure*}

In Figure \ref{fig:k-a}a, we plot all the 53 \atlas\ galaxies with \hi\ detections in the $K$-$A_{\rm F}$ plane,\footnote{We use $K$ instead of $K'$ for two reasons. First, $K$ is readily available from \hi\ spectra and does not rely on the sample-dependent fitting result in Equation \ref{eq:K0} and the different methods of deriving galaxy properties. Second, the difference between $K$ and $K'$ is small enough ($\sigma_{K_0}=0.02$ for xGASS). We have tested that using $K'$ does not affect our results.} using the same symbols and \hi\ morphological types as in \citet{Serra2012}. Galaxies from THINGS (dark red) and xGASS (gray) are also overlaid, along with contours enclosing 50\% and 90\% of the xGASS galaxies. ETGs with large, regularly rotating \hi\ disk ($D$-type) occupy a similar locus as xGASS and THINGS galaxies, consistent with their similarities in \hi\ distribution and kinematics. When \hi\ in ETGs is confined to a disk smaller than the stellar disk ($d$-type) or floating clouds ($c$-type), its distribution shows a much larger scatter in the plane. 

The most striking feature in Figure \ref{fig:k-a}a is that ETGs with unsettled \hi\ morphology and kinematics ($u$-type; where most \hi\ does not rotate regularly around the stellar body;  \citealt{Morganti2006, Serra2012, Lucero2013}) are confined within a narrow region around $K\sim0.08$ and $A_{\rm F}\gtrsim1.2$, suggesting a distinct parameter space for identifying unsettled \hi\ with high purity and completeness. 
We visually inspect the \hi\ moment maps of galaxies other than the $c$-type ones and find that two of them can be classified as unsettled (THINGS: NGC 3077; $D$-type: UGC 3960; see Appendix \ref{app:mom0}).
Furthermore, one $u$-type ETGs outside this region at $A_{\rm F}<1.1$ is likely a misclassified $d$-type \citep[NGC 7280;][]{Serra2012}. 
These findings lead us to conclude that external perturbations of \hi\ can increase $K$ by making the \hi\ unsettled, which is usually accompanied by high spectral asymmetry. In Figure \ref{fig:k-a}b, we propose a boundary to separate galaxies with normal (the remaining $D$-types and THINGS galaxies) and unsettled \hi\ disks using a linear Support Vector Machine, which maximizes the margin between the two classes in the $K$-$A_{\rm F}$ plane:
\begin{equation}
    A_{\rm F}>-2.9K+1.3. \label{eq:k-a}
\end{equation}
We caution, however, that this boundary is empirically derived from THINGS and \atlas\ galaxies, which are mostly massive and nearby. Care should be taken when applying it to galaxies at higher redshifts or with lower masses. It may also be contaminated by galaxies such as $d$-type and $c$-type ones (though these two types of ETGs are more \hi-poor; \citealt{Serra2012}), but is highly confident in excluding $D$-type galaxies.

\subsection{Effects of the missing \hi\ in interferometry data} \label{subsec:k-diff}

Interferometry observations like THINGS can filter out the \hi\ with large angular scales and miss those with low column densities \citep[e.g.,][]{Pisano2014,Pingel2018,Wang2023a}. This missing \hi\ component, which accounts for up to $\sim50\%$ in \hi\ mass for THINGS galaxies, is mostly diffuse and exhibits lagged rotation and higher velocity dispersion \citep{Wang2024a}. These properties of diffuse \hi\ lead to an increased fraction of \hi\ around the systematic velocities, thus increasing the value of $K$.

\begin{figure*}
    \centering
    \includegraphics[width=\linewidth]{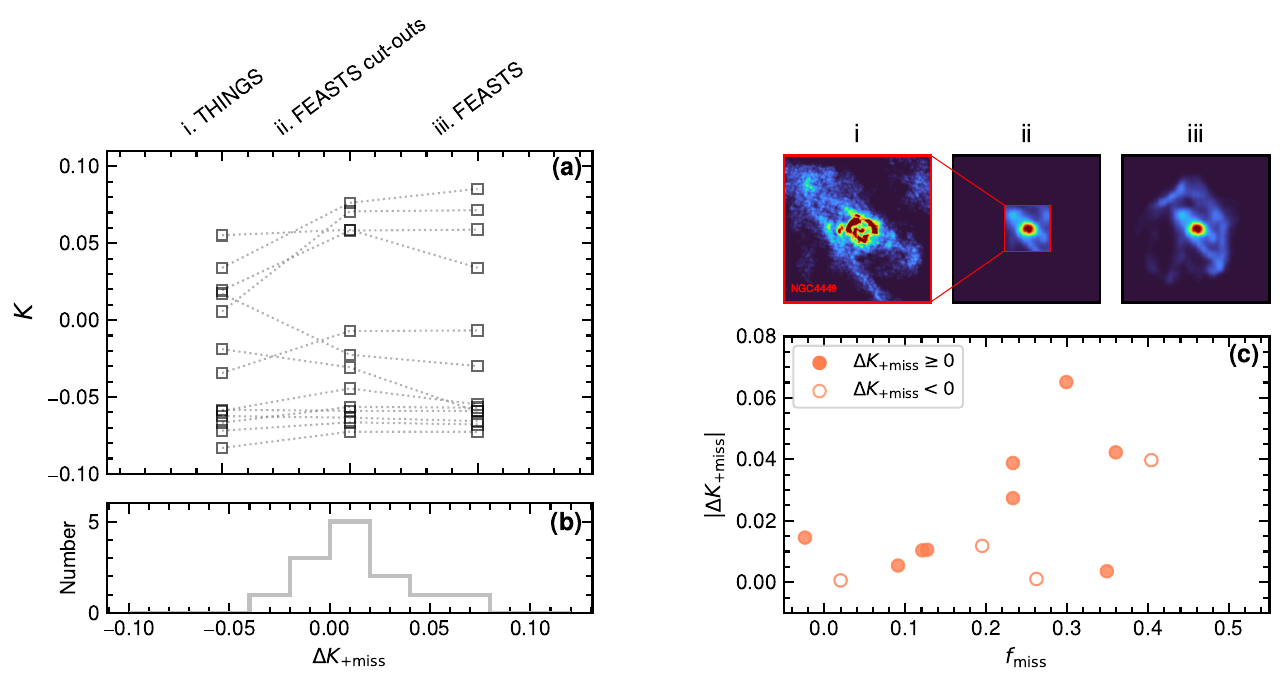}
    \caption{\textbf{(a)} Comparison of $K$ measured with different surveys for 13 galaxies observed with both THINGS and FEASTS. From left to right: (i) THINGS cubes; (ii) cut-outs of FEASTS cubes within the sky region of THINGS cubes; (iii) FEASTS full cubes. An example of the moment-0 maps of the three data sets is shown on the upper right panels. \textbf{(b)} Distribution of differences in $K$ between (i) and (ii) ($\Delta K_{\rm +miss}$). \textbf{(c)} Correlation between $|\Delta K_{\rm +miss}|$ and the mass fraction of missing \hi\ ($f_{\rm miss}$). The filled dots and open circles represent increased and decreased $K$ after adding \hi\ missed by THINGS, respectively.}
    \label{fig:k-miss}
\end{figure*}

We use 13 THINGS galaxies observed by the FEASTS project \citep{Wang2024a}. Because FEASTS observations have more complete coverage of a few \hi\ disks, we derive $K$ from the following data sets and compare them in Figure \ref{fig:k-miss}a: (i) THINGS cubes; (ii) cut-outs of FEASTS cubes within the sky region of THINGS cubes; (iii) FEASTS full cubes with more complete coverage on \hi\ disks for a few galaxies. We use NGC 4449 as an example to demonstrate the differences between these data sets. Comparing (i) and (ii), we find that the inclusion of missing \hi\ slightly increases $K$ by $\Delta K_{\rm +miss}=0.013\pm0.026$ (Figure \ref{fig:k-miss}b; $\Delta K_{\rm +miss}$ is defined as $K_{\rm (ii)}-K_{\rm (i)}$). This effect can be degenerated with, or partially contribute to the effect of spatial concentrated \hi\ (Section \ref{subsec:k-prof}). On the other hand, a comparison between (ii) and (iii) shows that \hi\ at the very large radii does not affect $K$ significantly. 
In Figure \ref{fig:k-miss}c, we present the relationship between the effect of missing \hi\ on $K$ and the mass fraction of missing \hi\ ($f_{\rm miss}$). As expected, the value of $K$ changes little for low $f_{\rm miss}$ ($<0.2$). At high $f_{\rm miss}$, however, $K$ exhibits diverse behaviors---it may increase, decrease, or stay relatively constant. This suggests that the specific distribution and kinematics of the missing \hi\ are important in determining $K$. Future efforts will be needed to enlarge the sample for a better understanding of the conditions driving the diversity. 

\subsection{How does $K$ reveal the evolution of PSBs?} \label{subsec:k-evo}

\begin{figure*}
    \centering
    \includegraphics[width=\linewidth]{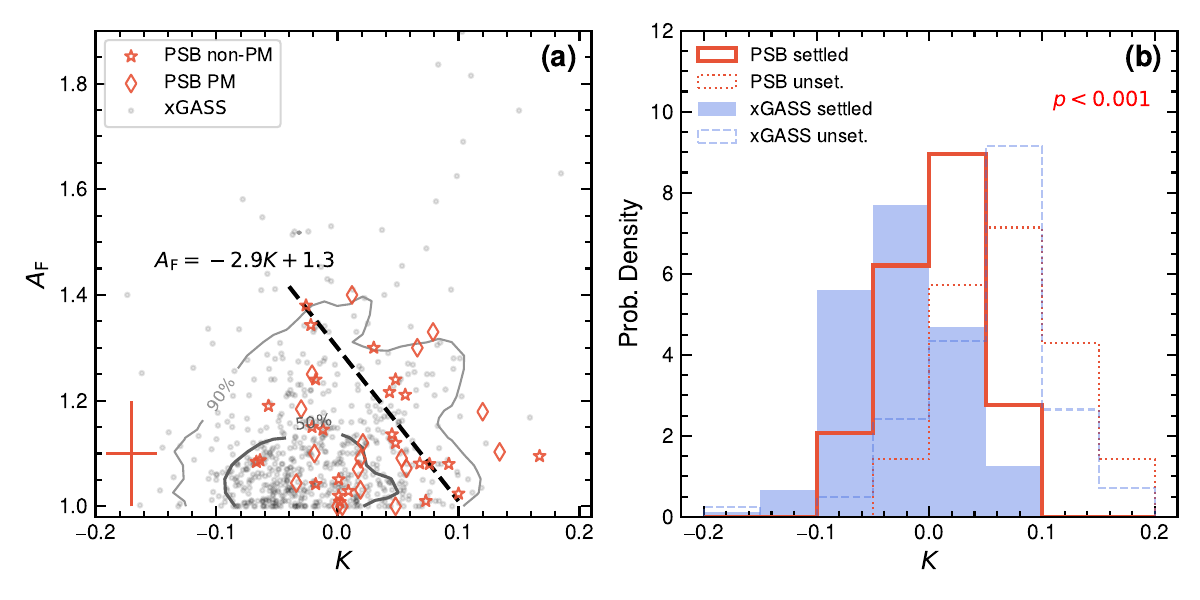}
    \caption{\textbf{(a)} Distributions of PSBs and xGASS galaxies on the $K$-$A_{\rm F}$ plane. Contours enclosing 50\% and 90\% xGASS galaxies are shown in gray, as in Figure \ref{fig:k-a}a. PSBs with and without post-merger (PM) features in their optical images are overlaid using red stars and red diamonds, respectively. The error bar in the lower left represents the typical measurement uncertainty of PSBs. \textbf{(b)} Similar to Figure \ref{fig:K-pdf}, except that galaxies defined by Equation \ref{eq:k-a} are separated from their parent samples.}
    \label{fig:k-a-psb}
\end{figure*}

Based on the previous discussions, we are now in a position to understand how the elevated $K$ values in PSBs are connected to their unique SFHs. As a first attempt, we examine the distribution of PSBs in the $K$-$A_{\rm F}$ plane (red symbols in Figure \ref{fig:k-a-psb}a). Less than one-third of the PSBs fall within the region associated with unsettled \hi\ (Equation \ref{eq:k-a}), while the remaining align with the distribution of normal \hi\ disks of xGASS galaxies and $D$-type ETGs. When we restrict the galaxies to those not classified as unsettled, PSBs still exhibit significantly higher $K$ values compared to xGASS galaxies (Figure \ref{fig:k-a-psb}b; the same holds for $K'$), with an average $K$ difference of 0.029. For these PSBs, the \hi\ is likely to be more spatially concentrated, supported by a higher level of turbulence, and may be accompanied by a low-density envelope diffusely distributed in the galaxy. Fig. \ref{fig:k-prof} shows that the concentrated \hi\ is likely distributed at $r<R_{\rm e}$, consistent with the concentrated CO emission at the central 1 kpc found in low-redshift PSBs \citep[e.g.,][]{Otter2022}. CO observations of our PSB sample will be presented in Rasmussen et al. (in prep).

Galaxy mergers are one of the main triggers of low-redshift PSBs \citep[e.g.,][]{Sazonova2021}, and they can significantly alter the gas kinematics. Through visual inspection of the optical images from Legacy Surveys \citep{Dey2019}, we identified 18 post-mergers (PMs) from our PSB sample (Appendix \ref{app:tab}), and the PM fraction ($42\%\pm10\%$) is consistent with previous studies \citep[e.g.,][]{Alatalo2016, Wilkinson2022, Ellison2024}. The PMs have a broad distribution in the $K$-$A_{\rm F}$ plane (red diamonds in Figure \ref{fig:k-a-psb}a), implying that while violent mergers destroy the \hi\ disks and leave the \hi\ unsettled, mild mergers can also remove angular momentum from the gas and lead to a concentrated gas distribution \citep[e.g.,][]{Blumenthal2018, Sparre2022}. Because the \hi\ disk extends further than the stellar disk in gas-rich galaxies, they are more easily disturbed by external forces and tidal features of \hi\ can survive longer than their stellar counterparts. This can explain why, above the threshold defined for unsettled \hi, only PSBs located furthest above the line are classified as PMs. 
\citet{Thorp2024} found that PMs tend to have more concentrated starbursts than their isolated counterparts. It would be interesting to see if the distribution of \hi\ in PSBs follows the same trend. Unfortunately, this is not possible here due to the limited sample size. Future progress can be made through both increasing the sample size and directly mapping the \hi\ distribution with interferometers.

\citetalias{Ellison2025} have demonstrated that PSBs are not devoid of \hi. Now we further show that the \hi\ central concentration and gas density are elevated in PSBs, in line with their CO properties. Under such conditions, intense star formation would be expected, which is indeed the case for PSBs in their recent past. From the similarity in the $K$ distributions between PSBs and starburst galaxies (Figure \ref{fig:K-pdf}b), we can infer that the elevated $K$ and the preceding starburst phase are induced by the same mechanisms such as mergers, and that the \hi\ concentration remains high even after star formation activity has declined. During this process, \hi\ is not significantly ejected from the PSBs by stellar or AGN feedback, as evidenced by the large remaining reservoirs found by \citetalias{Ellison2025}. Instead, it is stabilized against collapsing and forming stars, likely due to the heating processes \citep{Michalowski2024}, turbulence, or the formation of a spheroidal component during the starburst \citep{Martig2009}. For PSBs with the lowest $K$, it is possible that the \hi\ will settle into a cold disk and reignite star formation in these galaxies.

\section{Conclusion} \label{sec:conclusion}

The shape of integrated \hi\ spectra provides valuable information on the distribution and kinematics of neutral gas in galaxies. In this paper, we analyze the spectral concentration ($K$) of the largest and most complete sample of PSBs with deep \hi\ observations to date, and compare it with the representative xGASS sample. To understand the drivers of spectral concentration, we also construct integrated \hi\ spectra for galaxies with resolved \hi\ maps, including 34 late-type and dwarf galaxies in THINGS and 51 ETGs in \atlas. 
Several conclusions can be drawn based on our examinations of \hi\ spectral shape:

\begin{enumerate}
    \item PSBs exhibit significantly higher \hi\ spectral concentrations ($K$) compared to typical xGASS galaxies, with spectra more likely to be single-peaked rather than double-horned. This conclusion also holds for $K'$, which removes the dependence of $K$ on other related galaxy properties. The $K'$ distribution of PSBs is comparable to that of starburst galaxies in xGASS. 
    \item When corrected for inclination angles, $K$ shows a strong correlation with the spatial concentration of \hi\ in THINGS galaxies. This correlation is strongest within the effective radius ($R_{\rm in}\sim 0.7R_{\rm e}$) and becomes insignificant in the outer disk at $R_{\rm in}\gtrsim 2.5R_{\rm e}$.
    \item Non-axisymmetric structures introduce random variations in $K$, particularly through the $m=1$ (lopsidedness) and $m=2$ (dipole) Fourier modes.
    \item Galaxies with unsettled \hi\ distributions display both high $K$ and high spectral asymmetry ($A_{\rm F}$). This allows for the identification of such galaxies in the $K$-$A_{\rm F}$ plane.
    \item Diffuse \hi\ in galaxies can systematically increase $K$ by $\sim0.01$, as it is generally more dispersion dominated than dense \hi. For individual galaxies, however, the effect varies depending on the spatial distribution and kinematics of diffuse \hi.
\end{enumerate}

Based on these results, we conclude that the elevated $K$ and the quenching of star formation in PSBs likely result from a combination of mechanisms. In around a quarter of the PSBs, star formation ceases mainly because the gas is not in the right position (unsettled). While in the remaining PSBs, the centrally concentrated or diffuse \hi\ suggests that the cold gas is not in the right state to form stars. A significant fraction of post-mergers are identified in both categories, which implies that galaxy interactions play a key role in increasing $K$ by driving gas inflows or disrupting the \hi\ disk. A specific study of how mergers affect the \hi\ spectral shape will be useful, and considering a large sample of spatially resolved \hi\ data is necessary in the future \citep[e.g., WALLABY;][]{Koribalski2020}.

\section*{}

We thank the anonymous referee for valuable comments that improved this manuscript. We thank Vivienne Wild for kindly providing the SDSS spectra fitting results for PSBs. JW thanks support of research grants from  Ministry of Science and Technology of the People's Republic of China (NO. 2022YFA1602902), National Science Foundation of China (NO. 12233001), and the China Manned Space Project (No. CMS-CSST-2025-A08). SLE acknowledges the receipt of an NSERC Discovery Grant.

This work made use of the data from FAST (Five-hundred-meter Aperture Spherical radio Telescope) (https://cstr.cn/31116.02.FAST). FAST is a Chinese national mega-science facility, operated by National Astronomical Observatories, Chinese Academy of Sciences. This work is based on data obtained with the Westerbork Synthesis Radio Telescope (WSRT). The WSRT is operated by ASTRON (Netherlands Institute for Radio Astronomy) with support from the Netherlands Foundation for Scientific Research (NWO). The National Radio Astronomy Observatory is a facility of the U.S. National Science Foundation operated under cooperative agreement by Associated Universities, Inc.

The DESI Legacy Imaging Surveys consist of three individual and complementary projects: the Dark Energy Camera Legacy Survey (DECaLS), the Beijing-Arizona Sky Survey (BASS), and the Mayall z-band Legacy Survey (MzLS). DECaLS, BASS and MzLS together include data obtained, respectively, at the Blanco telescope, Cerro Tololo Inter-American Observatory, NSF’s NOIRLab; the Bok telescope, Steward Observatory, University of Arizona; and the Mayall telescope, Kitt Peak National Observatory, NOIRLab. NOIRLab is operated by the Association of Universities for Research in Astronomy (AURA) under a cooperative agreement with the National Science Foundation. Pipeline processing and analyses of the data were supported by NOIRLab and the Lawrence Berkeley National Laboratory (LBNL). Legacy Surveys also uses data products from the Near-Earth Object Wide-field Infrared Survey Explorer (NEOWISE), a project of the Jet Propulsion Laboratory/California Institute of Technology, funded by the National Aeronautics and Space Administration. Legacy Surveys was supported by: the Director, Office of Science, Office of High Energy Physics of the U.S. Department of Energy; the National Energy Research Scientific Computing Center, a DOE Office of Science User Facility; the U.S. National Science Foundation, Division of Astronomical Sciences; the National Astronomical Observatories of China, the Chinese Academy of Sciences and the Chinese National Natural Science Foundation. LBNL is managed by the Regents of the University of California under contract to the U.S. Department of Energy. The complete acknowledgments can be found at https://www.legacysurvey.org/acknowledgment/.

\appendix

\section{Dependencies of $K$} \label{app:k-dep}
\begin{figure*}
    \centering
    \includegraphics[width=\linewidth]{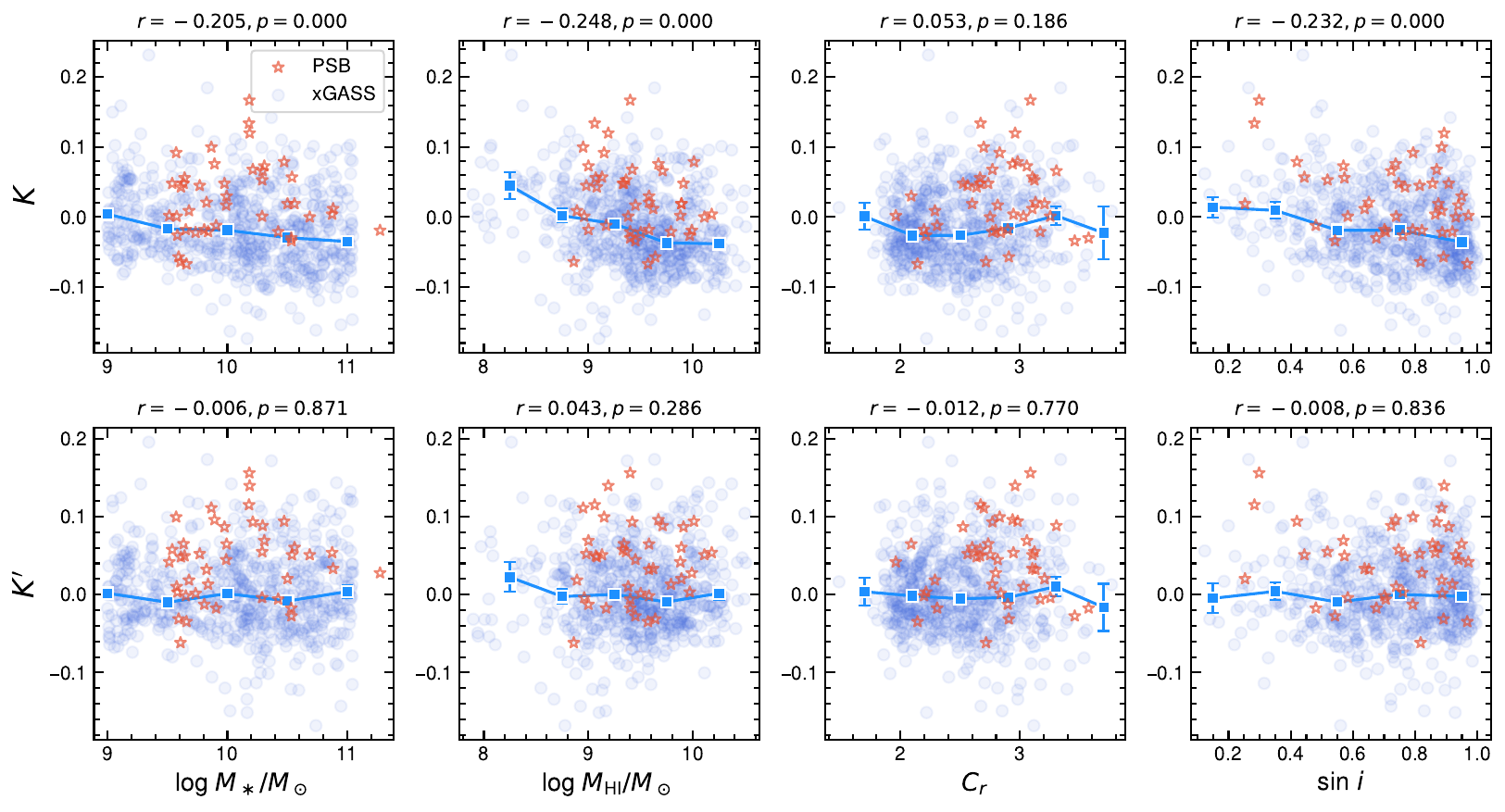}
    \caption{Dependence of $K$ (upper) and $K'$ (lower) on stellar mass, \hi\ mass, optical concentration, and inclination angle. Blue dots and red stars represent xGASS galaxies and PSBs, respectively. The Pearson-$r$ coefficient and the $p$-value for xGASS are shown at the top of each panel. The median trends of xGASS are plotted as solid blue lines.}
    \label{fig:k-dep}
\end{figure*}

In Section \ref{sec:result}, we mentioned that $K$ is dependent on other galaxy properties, and removing these dependencies allows a more adequate comparison between PSBs and xGASS. In the upper panels of Figure \ref{fig:k-dep}, we show the correlation between $K$ and $M_\ast$, $M_{\rm HI}$, $C_r$, and $\sin i$. The lower panels demonstrate that after performing the linear fitting in Equation \ref{eq:K0}, the $K$-excess ($K'$) exhibits no significant correlation with these galaxy properties anymore (all $|r|<0.1$, $p>0.1$). This suggests that $K'$ is a more robust metric for comparing PSBs to the xGASS sample.

\section{Dealing with THINGS and \atlas\ data} \label{app:int}

Here we describe the detailed procedures for the mock tests discussed in Section \ref{subsec:k-naxis} and the method to obtain the integrated \hi\ spectra for \atlas\ galaxies (Section \ref{subsec:k-a}). 

\subsection{THINGS}

The integrated \hi\ spectrum of THINGS are generated by combining the moment-0, moment-1, and moment-2 maps, assuming a Gaussian line profile at each spaxel. For further analysis in Section \ref{subsec:k-naxis}, we deproject the observed \hi\ surface density ($\Sigma_{\rm HI}$) maps onto the disk plane using the geometry parameters (center, P.A., $i$) listed in \citet{Walter2008} and assuming that the \hi\ disk is infinitely thin. Then, we perform azimuthal Fourier decomposition on a series of 1-pixel wide annulus to obtain the radial profiles of amplitudes $A_{\rm m}(r)$ and phase angles $\phi_{\rm m}(r)$:
\begin{equation}
    \Sigma_{\rm HI}(r,\phi)=\Sigma_{0}(r)\left\{1+\sum_{\rm m}A_{\rm m}(r)\cos\left[ {\rm m}\phi-\phi_{\rm m}(r)\right]\right\}. \label{eq:fft}
\end{equation}

To rotate the moment-0 maps in the disk plane, we simply add constant shifts to the phase angles at each radius and reconstruct the deprojected moment-0 maps using Equation \ref{eq:fft}, which are then projected into the observed frame. The rotation angle ($\varphi_{\rm rot}$) ranges from $0^\circ$ to $350^\circ$, with increments of $10^\circ$. For each rotated moment-0 map, we generate the corresponding integrated spectrum and measure the spectral concentration $K$. The $\Delta K$ is calculated by subtracting the $K$ value of the $\varphi_{\rm rot}=0^\circ$ spectrum from that of the target spectrum. Varying the Fourier amplitudes, as in Figure \ref{fig:k-naxis}, will result in pixels with negative fluxes in the reconstructed maps. In such cases, we set the values of these pixels to zero before generating the integrated \hi\ spectra. The deprojected moment-0 maps are reconstructed using Fourier modes with $m<20$. The threshold is chosen as a balance between the robustness of the reconstruction and the computational speed.

\subsection{\atlas}
We use the fully reduced WRST cubes and moment-0 maps\footnote{\url{https://www-astro.physics.ox.ac.uk/atlas3d/}} to derive the integrated \hi\ spectra of \atlas\ galaxies. For each galaxy, its spectrum is extracted from the cube by summing up the spectra at spaxels within the masks of the moment-0 maps. In three galaxies (NGC 680, NGC 3626, NGC 5103), the moment-0 maps of our targets are blended with their neighbors, and we perform a 2-D source de-blending using \texttt{Photutils} \citep{Bradley2022} to separate the spaxels using the moment-0 maps. A 3-D de-blending method will not improve the results significantly in these cases, as the \hi\ disks are well separated in the projected sky plane \citep{Huang2025}. 

\section{Moment-0 maps of NGC 3077 and UGC 3960} \label{app:mom0}

\begin{figure*}
    \centering
    \includegraphics[width=0.9\linewidth]{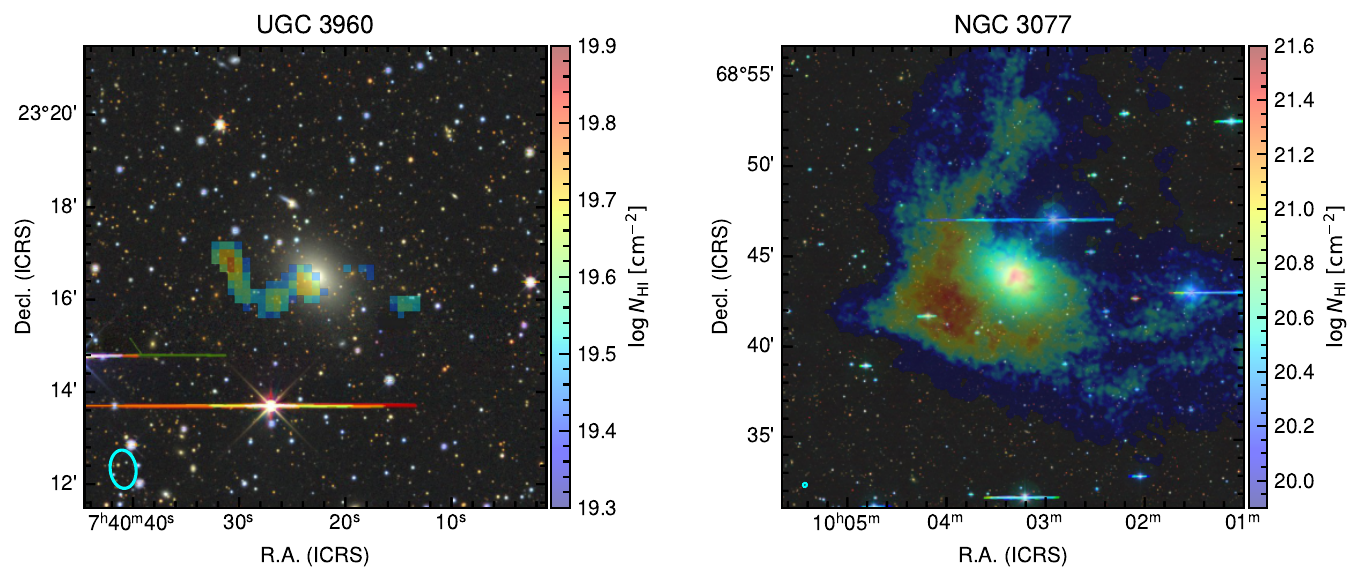}
    \caption{The \hi\ moment-0 maps of UGC 3960 (left; \atlas) and NGC 3077 (right; THINGS). The background images are from the Legacy Surveys \citep{Dey2019}. The cyan circles in the lower left indicate the synthesized beam.}
    \label{fig:atlas}
\end{figure*}

Figure \ref{fig:atlas} shows the \hi\ moment-0 maps of two galaxies classified as unsettled: (1) NGC 3077, the $D$-type \atlas\ galaxy closest to the locus of $u$-types in the $K$-$A_{\rm F}$ plane; and (2) UGC 3960, the THINGS galaxy closest to the locus of $u$-types in the $K$-$A_{\rm F}$ plane (Figure \ref{fig:k-a}a).

\newpage
\section{Properties of the PSB sample} \label{app:tab}

Table \ref{tab:psb} presents the $K$ values and morphology flags for the PSB sample. The flags are obtained through visually inspecting the optical images from Legacy Surveys \citep{Dey2019}.

\begin{deluxetable*}{cccccccccc}
\tablecaption{Properties of 43 PSBs in the final PSB sample.} \label{tab:psb}
\tablewidth{\textwidth}
\decimalcolnumbers
\tabletypesize{\small}
\tablehead{SDSS objID & R.A. & Decl. & $z$ & $\log M_\ast$ & $\log M_{\rm HI}$ & PM flag & $K$ & $K'$ & $\sigma_K$ \\
 & (deg) & (deg) &  & ($M_\odot$) & ($M_\odot$) & & & & }
\startdata
587722984440463382 & 216.5541 & 0.8606 & 0.03186 & 10.18 & 9.06 & T & 0.134 & 0.115 & 0.048 \\
587725489988960505 & 257.1393 & 57.4767 & 0.02961 & 9.61 & 8.86 & F & -0.064 & -0.062 & 0.042 \\
587726032256630848 & 198.4683 & 2.1326 & 0.03026 & 10.53 & 9.40 & T & -0.030 & -0.017 & 0.028 \\
587729652890206755 & 256.4333 & 31.4138 & 0.03480 & 9.81 & 9.00 & F & -0.018 & -0.013 & 0.020 \\
587730023333232703 & 230.5217 & 5.8549 & 0.03564 & 10.50 & 9.89 & T & 0.019 & 0.020 & 0.018 \\
587730773885059094 & 351.2256 & 14.2152 & 0.02563 & 9.99 & 9.45 & T & 0.017 & 0.045 & 0.023 \\
587730847963545655 & 318.5023 & 0.5351 & 0.02692 & 10.18 & 9.40 & F & 0.167 & 0.156 & 0.054 \\
587731174382502294 & 319.9507 & 0.6727 & 0.03442 & 9.61 & 9.08 & F & 0.043 & 0.048 & 0.057 \\
587731186735186207 & 347.2619 & 0.2669 & 0.03252 & 10.19 & 9.19 & T & 0.120 & 0.140 & 0.015 \\
587731886809808959 & 123.8574 & 37.3405 & 0.03975 & 9.89 & 9.67 & F & 0.076 & 0.096 & 0.048 \\
587732580982521898 & 169.7818 & 58.0540 & 0.03260 & 10.54 & 9.13 & T & 0.057 & 0.051 & 0.023 \\
587734622705811566 & 131.1576 & 32.9064 & 0.03154 & 10.31 & 9.00 & F & 0.073 & 0.069 & 0.018 \\
587734891673026666 & 165.3640 & 8.4206 & 0.03061 & 9.84 & 9.79 & F & -0.021 & 0.013 & 0.020 \\
587732771584671859 & 164.5907 & 9.4539 & 0.03359 & 9.77 & 8.98 & F & 0.045 & 0.052 & 0.037 \\
587735347486457882 & 149.9206 & 11.5331 & 0.03669 & 10.42 & 9.40 & T & -0.021 & -0.006 & 0.037 \\
587735664773431424 & 226.1544 & 48.7388 & 0.03610 & 10.53 & 9.45 & T & -0.034 & -0.027 & 0.009 \\
587736584980463705 & 251.1281 & 19.9408 & 0.02300 & 10.29 & 9.36 & T & 0.053 & 0.055 & 0.037 \\
587736541491233170 & 237.7184 & 5.3269 & 0.02603 & 9.66 & 9.57 & F & -0.067 & -0.035 & 0.010 \\
587736808838594663 & 207.9975 & 13.9675 & 0.03669 & 11.27 & 9.99 & T & -0.019 & 0.028 & 0.021 \\
587738615415373880 & 170.9459 & 35.4423 & 0.03407 & 10.29 & 8.96 & F & 0.001 & -0.004 & 0.017 \\
587739130805420150 & 206.4464 & 34.4932 & 0.03479 & 9.54 & 9.33 & F & 0.048 & 0.058 & 0.008 \\
587739504477077652 & 204.0172 & 30.1411 & 0.02593 & 9.64 & 9.07 & F & 0.056 & 0.050 & 0.016 \\
587739646208770144 & 168.9029 & 30.4228 & 0.02788 & 9.65 & 9.58 & F & -0.018 & -0.002 & 0.034 \\
587739828743962776 & 228.0094 & 21.2982 & 0.01578 & 10.69 & 10.11 & T & -0.000 & 0.051 & 0.015 \\
587739845393186912 & 240.2146 & 15.1513 & 0.03396 & 10.47 & 10.01 & T & 0.079 & 0.094 & 0.032 \\
587741532777152653 & 149.7504 & 25.1031 & 0.02209 & 9.57 & 9.15 & F & 0.092 & 0.099 & 0.043 \\
587741721214386184 & 197.9218 & 26.3901 & 0.03811 & 9.98 & 9.91 & T & 0.048 & 0.087 & 0.033 \\
587742010042744920 & 121.1948 & 10.7783 & 0.03526 & 10.87 & 10.18 & T & 0.004 & 0.054 & 0.017 \\
587742189908983921 & 194.5416 & 24.3489 & 0.02267 & 9.58 & 9.38 & F & 0.001 & 0.010 & 0.021 \\
587742576459251895 & 225.3402 & 15.2499 & 0.03537 & 9.91 & 9.17 & F & -0.012 & -0.017 & 0.012 \\
587742551762796902 & 238.8105 & 12.9163 & 0.03291 & 9.58 & 9.97 & F & -0.026 & 0.003 & 0.021 \\
587742616170266870 & 235.7542 & 16.9874 & 0.03141 & 9.87 & 8.95 & F & 0.100 & 0.111 & 0.018 \\
587742627998531885 & 246.2755 & 8.3820 & 0.03511 & 10.29 & 9.71 & T & 0.066 & 0.088 & 0.033 \\
587742628523475193 & 219.3072 & 14.6651 & 0.03790 & 10.88 & 9.62 & T & 0.012 & 0.033 & 0.042 \\
587745244159082657 & 136.4664 & 13.7175 & 0.02735 & 9.68 & 8.89 & F & 0.009 & 0.019 & 0.037 \\
588015508189872263 & 346.9322 & -0.8385 & 0.03253 & 9.52 & 9.88 & F & 0.002 & 0.042 & 0.003 \\
588016891172618376 & 147.9371 & 35.6221 & 0.02705 & 10.56 & 9.89 & T & 0.019 & 0.060 & 0.005 \\
588017627780087861 & 227.2295 & 37.5583 & 0.02906 & 10.21 & 9.42 & F & 0.068 & 0.093 & 0.010 \\
588017724947759115 & 208.8698 & 6.5964 & 0.02408 & 9.83 & 9.58 & T & 0.021 & 0.032 & 0.019 \\
588848898848784582 & 202.5796 & -0.8707 & 0.03760 & 9.73 & 9.50 & F & -0.022 & 0.001 & 0.041 \\
588017992299380883 & 202.2052 & 10.3817 & 0.02310 & 9.59 & 9.64 & F & -0.057 & -0.031 & 0.017 \\
588023046941245549 & 139.1146 & 19.9205 & 0.02598 & 9.99 & 9.58 & F & 0.030 & 0.065 & 0.021 \\
588298664112881826 & 192.6636 & 47.9343 & 0.02911 & 9.63 & 9.34 & F & 0.048 & 0.065 & 0.027 \\
\enddata
\tablecomments{Columns: (1) SDSS DR7 ObjID. (2) Right ascension. (3) Declination. (4) Redshift from SDSS DR7 optical spectrum. (5) Stellar mass from the MPA-JHU catalogs. (6) \hi\ mass from \citetalias{Ellison2025}. (7) Visual identification of post-mergers. (8) Concentration of the \hi\ spectra. (9) The $K$-excess, defined as $K-K_0$ (Section \ref{sec:result}). (10) Uncertainty of $K$.}
\end{deluxetable*}

\newpage
\bibliography{Huang+25_psb}{}
\bibliographystyle{aasjournal}

\end{CJK*}
\end{document}